\documentstyle[12pt,psfig]{article}
\newcommand{\NP}[1]{ Nucl.\ Phys.\ {\bf #1}}

\newcommand{\PR}[1]{Phys.\ Rev.\ {\bf #1}}
\newcommand{\PRL}[1]{ Phys.\ Rev.\ Lett.\ {\bf #1}}

\begin{document}
\begin{flushright}
SLAC-PUB-8033\\
June 1999
\end{flushright}
\vfill
\begin{center}
{\LARGE {Measuring the QCD Gell Mann-Low $\Psi$-function}
\footnote{Research partially supported
by the Department of Energy under contract DE-AC03-76SF00515
and the Spanish CICYT under contracts AEN-96-1673, AEN97-1693.
}}\\[2ex]
S. J. Brodsky$^1$, C. Merino$^2$, and J. R. Pel\'aez$^3$
\footnote{E-mail:pelaez@eucmax.sim.ucm.es}\\
{\small\em $^1$ Stanford Linear Accelerator Center}\\
{\small\em Stanford University,
Stanford, California 94309. U.S.A.}

{\small\em $^2$ Departamento de F\'{\i}sica de Part\'{\i}culas.}\\
{\small\em Universidade de Santiago de Compostela,
15706 Santiago de Compostela, Spain.}

{\small\em $^3$ Departamento de F\'{\i}sica Te\'orica.}\\
{\small\em Universidad Complutense de Madrid,
28040 Madrid, Spain.}
\end{center}

\begin{center}
Abstract
\end{center}
We present a general method for extracting the  Gell
Mann-Low logarithmic derivative  of an effective charge of an
observable directly from data as a mean for empirically verifying the
universal terms of the QCD $\beta$-function. Our method avoids
the biases implicit in fitting to QCD-motivated forms as well as the
interpolation errors introduced by constructing
derivatives from discrete data. We also derive relations between
moments of effective charges as new tests of perturbative QCD.
\vfill
\centerline{Submitted to Physical Review D}
\vfill
\newpage

\section{Introduction}

An effective charge~\cite{effcha} encodes the entire perturbative
correction of a QCD observable; for example, the ratio of  $e^+ e^-
\gamma^* \to {\rm hadrons}$ annihilation  to  muon pair
cross sections  can be written
\begin{equation}
R_{e^+e^-}(s)\equiv\frac{\sigma(e^+e^-\rightarrow {\rm hadrons})}{
\sigma(e^+e^-\rightarrow\mu^+\mu^-)}
=R_{e^+e^-}^0(s)\left(1+\frac{\alpha_R(\sqrt s)}{\pi}\right),
\label{Ree}
\end{equation}
where $R^0_{e^+ e^-}$ is the prediction at Born level.
More generally, the effective charge $\alpha_A(Q)$ is defined as
the entire QCD radiative contribution to an observable ${\cal
O}_A(Q)$~\cite{effcha}:
\begin{equation}
{\cal O}_A(\Lambda)={\cal
O}_A^0\left(\delta_A+\frac{\alpha_A(\Lambda)}{\pi}\right),
\label{RA}
\end{equation}
where $\delta_A$ is the zeroth order QCD prediction (i.e., the parton model),
and $\alpha_A(\Lambda)/\pi$ is the entire QCD correction. Note that
$\delta_A=0$ or $1$ depending on whether the observable A exists at zeroth
order.
Important examples with $\delta_A=1$ are the $e^+ e^-$ annihilation
cross-section ratio
and the $\tau$ lepton's hadronic decay ratio,
\begin{equation}
R_{\tau} \equiv\frac{\Gamma(\tau^-\rightarrow{\nu_{\tau} +
{\rm hadrons}})}{\Gamma(\tau^-\rightarrow{\nu_{\tau}e^-\bar{\nu_e}})}
= R_\tau^0\left(1+\frac{\alpha_\tau(m_\tau)}{\pi}\right).
\end{equation}
In contrast, the effective charge $\alpha_V(Q)$ defined from the
static heavy quark potential and the effective charge
$\alpha_{>2\,\hbox{jets}}$ defined from $e^+e^-$ annihilation into more
than two jets,
$\sigma_{>2\,\hbox{jets}}$, have $\delta_A=0$.

One can define
effective charges for virtually any quantity calculable in
perturbative QCD; e.g., moments of structure functions, ratios of form
factors, jet observables, and the effective potential between massive
quarks. In the case of decay constants of the $Z$ or the $\tau$, the mass
of the decaying system serves as the physical scale in the effective
charge. In the case of multi-scale observables, such as
the two-jet fraction in
$e^+ e^-$ annihilation, the arguments of the effective coupling
$\alpha_{2 jet}(s, y)$  correspond to the overall available energy and
characteristic kinematical jet mass fraction. Effective charges are {\em
defined} in terms of observables and, as such, are renormalization-scheme
and renormalization-scale independent.

The scale $Q$ which enters a given effective charge corresponds to its
physical momentum scale.   The total derivative of each
effective charge $\alpha_A(Q)$  with respect to the
logarithm of its physical scale is given by the Gell Mann-Low function:

\begin{equation}
 \Psi_A [\alpha_A(Q,m), Q/m] \equiv {d \alpha_A(Q,m) \over d \log Q} ,
\end{equation}
where the functional dependence of $\Psi_A$ is specific to the
effective charge $\alpha_A$.   Here $m$ refers to the quark's pole
mass.  The pole mass is universal in that it does not depend on the
choice of effective charge.   It should be
emphasized that the Gell Mann-Low $\Psi$ function is a property of a
physical quantity, and it is thus independent of conventions such as the
renormalization procedure and the choice of renormalization scale.

A central feature of quantum chromodynamics  is
asymptotic freedom; i.e., the monotonic decrease of the QCD coupling
$\alpha_A(Q^2)$ at large spacelike scales.  The empirical test of
asymptotic freedom is the verification of the negative sign of the Gell
Mann-Low function at large momentum transfer, a feature which must
in fact be true for any effective charge.

In perturbation theory,

\begin{equation}
\Psi_A = - \Psi_A^{\{0\}} {\alpha_A^2\over \pi}
          - \Psi_A^{\{1\}}{\alpha_A^3\over \pi^2}
          - \Psi_A^{\{2\}} {\alpha_A^4\over \pi^3}   + \cdots
        \end{equation}

At large scales $Q^2 >> m^2$, where the quarks can be treated as
massless, the first two terms are universal~\cite{t'Hooft} and basically 
given by the first two terms of the usual QCD $\beta$ function for
$N_C=3$
\begin{eqnarray}
\Psi_A^{\{0\}}&=&\frac{\beta_0}{2} = \frac{11}{2}\,  - \frac{1}{3}\,
   N_{F,A}^{\{0\}},\nonumber\\
\Psi_A^{\{1\}}&=&\frac{\beta_1}{8}   =
    \frac{51}{4}\, + \frac{19}{12}\, N_{F,A}^{\{1\}}.
\end{eqnarray}
Unlike the
$\beta$-function which controls the renormalization scale dependence of
bare couplings such as $\alpha_{\overline MS}(\mu)$, the
$\psi$ function is analytic in $Q^2/m^2$. In the case of the $\alpha_V$
scheme, the effective charge defined from the heavy quark potential, the
functional dependence of
$N_{F,V}(Q^2/m^2)$ is known to two loops~\cite{BGMR}.

The purpose of this paper is to develop an accurate method  for
extracting  the Gell Mann-Low function from measurements of an effective
charge in a manner which avoids the biases and uncertainties present
either in a standard fit or in numerical differentiation of
the data.  We will show  that one can indeed obtain
strong constraints on
$\Psi_A^{\{0\}}$ and
$\Psi_A^{\{1\}}$ from generalized moments of the measured quantities
which define the effective charge.
We find that the weight function $f(\xi)$ which
defines the effective charge
$\alpha_{Af}(\Lambda)$ from an integral of the effective charge
$\alpha_{A}(Q)$ can be chosen to produce maximum sensitivity to the
Gell-Mann Low function. As an example we will apply the method to the
$e^+e^-$ annihilation into more than two jets. Clearly one could also
extract the Gell Mann-Low function directly from a fit to the data, but
the fact that we are dealing with a logarithmic derivative introduces
large uncertainties \cite{Neubert}. Our results minimize some of these
uncertainties. In addition, our analysis provides a new class of
commensurate relations  between observables which are devoid of
renormalization scheme and scale artifacts.

One can define generalized effective charges from moments of the
observables. The classic example is $\alpha_{\tau}(\Lambda)$ where
$\Lambda$ is the generalization of the lepton mass.  The relevant point is
that
$R_\tau$ can be written as an integral of $R_{e^+e^-}$
\cite{Braaten}, as follows:
\begin{equation}
R_{\tau}(\Lambda^2)=\frac{2}{\sum_f q_f^2}\int_0^{\Lambda^2}\,
\frac{ds}{\Lambda^2}
\left( 1-\frac{s}{\Lambda^2}\right)^2\left( 1+\frac{2s}{\Lambda^2}\right)
R_{e^+e^-}(s),
\label{Rtauk}
\end{equation}
where $q_f$ are the quark charges. As a consequence of the mean value
theorem, the associated effective charges are related by a scale shift
\begin{equation}
\alpha_\tau(\Lambda)=\alpha_R(\sqrt s = \Lambda_\tau),
\label{CSRtau}
\end{equation}
The ratio of scales $\Lambda_{\tau}/\Lambda$   in
principle is predicted by QCD \cite{CSR}: The prediction at NLO is
\cite{CSR}
\begin{equation}
{\Lambda_{\tau} \over \Lambda} = \exp\left[ -{19\over 24} - {169\over 128}
{\alpha_R(\Lambda_\tau)\over \pi} + \cdots\right].
\label{lambdatau}
\end{equation}
Such relations between observables are called commensurate
scale relations (CSR)~\cite{CSR}.

The relation between $R_\tau$ and $R_{e^+e^-}$ suggests
that we can obtain additional useful effective charges by changing
the functional weight appearing in the integrand.
Indeed it has been shown \cite{We} that, starting
from any given observable ${\cal O}_A$
we can obtain new effective
charges $\alpha_{A\,f}$ by
constructing the following quantity
\begin{equation}
{\cal O}_{A\,f}(\Lambda) =
C\int_{\Lambda^2_1(\Lambda)}^{\Lambda^2_2(\Lambda)}\,
\frac{ds}{\Lambda^2} f\left(\frac{\sqrt{s}}{\Lambda}\right)
{\cal O}_A(\sqrt{s}),
\label{RAf}
\end{equation}
where $C$ is a constant and
$f(\xi)$ is a positive arbitrary integrable function. In order for 
${\cal O}_{A\,f}$ to define an effective charge $\alpha_{A\,f}$ through
\begin{equation}
{\cal O}_{A\,f}(\Lambda)={\cal O}^0_{A\,f}\left(\delta_A
+\frac{\alpha_{A\,f}(\Lambda)}{\pi}\right),
\label{alphaandO}
\end{equation}
it is necessary that $\Lambda_1(\Lambda)=\lambda_1\Lambda$ 
and $\Lambda_2(\Lambda)=\lambda_2\Lambda$, with both $\lambda_1$ 
and $\lambda_2$ constant.
Then, by the mean value theorem, $\alpha_{A\,f}$ is related again to
$\alpha_A$ by a scale shift
\begin{equation}
\alpha_{A\,f}(\Lambda)=\alpha_A(\Lambda_{Af}),
\label{alfaAk}
\end{equation}
with $\Lambda_1<\Lambda_{Af}<\Lambda_2$. An important
observation \cite{We} is that PQCD predicts
$\lambda_{Af}=\Lambda_{Af}/\Lambda$ to leading
twist.
If we ignore quark masses so that the two first
coefficients of the Gell-Mann Low function are constant, one has
\begin{eqnarray}
&&\frac{\alpha_A(\sqrt{s})}{\pi}= \frac{\alpha_A(\Lambda)}{\pi} -
\frac{\Psi_0}{2} \ln{\left(\frac{s}{\Lambda^2}\right)}
\left(\frac{\alpha_A(\Lambda)}{\pi}\right)^2  + \label{aR2order} \\
\nonumber
&&\quad + \frac{1}{4}\left[
\Psi_0^2 \ln^2\left(\frac{s}{\Lambda^2}\right)-
2\,\Psi_1 \ln\left(\frac{s}{\Lambda^2}\right)\right]
\left(\frac{\alpha_A(\Lambda)}{\pi}\right)^3
\ldots
\end{eqnarray}
If we now use  eqs.(\ref{RAf}) and (\ref{alphaandO}), we find \cite{We}
\begin{eqnarray}
\frac{\alpha_{Af}(\Lambda)}{\pi} &=& \frac{\alpha_A(\Lambda)}{\pi} -
\frac{\Psi_0}{2}\frac{I_{1f}}{I_{0f}}
\left(\frac{\alpha_A(\Lambda)}{\pi}\right)^2 \nonumber \\
&+& \frac{1}{4}\left[\Psi_0^2 \frac{I_{2f}}{I_{0f}}-
2\,\Psi_1 \frac{I_{1f}}{I_{0f}}\right]
\left(\frac{\alpha_A(\Lambda)}{\pi}\right)^3
\ldots,
\label{aRf2order}
\end{eqnarray}
where
$I_{lf} = \int^{\lambda^2_2}_{\lambda^2_1}f(\xi)(\ln{\xi^2})^ld\,\xi^2$
is independent of the choices of observable $A$ and scale $\Lambda$,
but only 
provided that $\Lambda_1(\Lambda)=\lambda_1\Lambda$ and 
$\Lambda_2(\Lambda)=\lambda_2\Lambda$.
Replacing $s$ by $\Lambda^2_{A\, f}$ in eq. (\ref{aR2order}) and comparing
with eq. (\ref{aRf2order}), we find
\begin{equation}
\lambda_{Af} = \exp\left\{\frac{I_{1f}}{2I_{0f}} +
\frac{\Psi_0}{4}\left[\left(\frac{I_{1f}}{I_{0f}}\right)^2
- \frac{I_{2f}}{I_{0f}}
\right]\frac{\alpha_A(\Lambda)}{\pi}\; ...\right\}.
\label{sstar}
\end{equation}
In general
the commensurate scale relation will have the
following expansion
\begin{equation}
\ln\lambda_{Af}(\Lambda)=\sum_{n=0}^\infty a_f^{(n)}
\left(\frac{\alpha_A(\Lambda)}{\pi}\right)^n,
\label{lexpand}
\end{equation}
where the first three coefficients are independent of $A$. Note that the above
formulae are only valid inside regions of constant $N_F$ and sufficiently
apart from quark thresholds. If we include the mass dependence, the
effective charges, by the mean value theorem, are still related by a scale
shift, although it cannot be written in the simple form of 
eq. (\ref{sstar}). Indeed, even
the lowest order of
$\lambda_{A\,f}$ would have a small dependence on the energy and
the effective number of flavors appearing in $\Psi_0$.

\section{Obtaining the Gell Mann-Low function  directly from observables}

The main practical obstacle in determining the Gell Mann-Low function
from experiment is that it is a {\em logarithmic} derivative.
One can try to obtain the value of the parameters of the $\Psi$
function from a direct fit to the data using the QCD forms, but any
approximation to the derivative
of the experimental results implicitly requires extrapolation or
interpolation of the data.
In order to observe a significant variation of the effective charge
$\alpha_A$ one needs to compare two vastly separated scales.
This is illustrated in Fig.1. However, to approximate
$\Psi(\sqrt{s})\simeq \Delta \alpha_A(\sqrt{s})/(\Delta \ln \sqrt{s})$
with a huge separation between $\sqrt{s}$ and $\sqrt{s'}$ is not very accurate
since then the value for $\Delta \alpha_A/\Delta \ln \sqrt{s}$ 
is the slope of the $Q$ straight line in Fig. 1
instead of that of $P$, which gives  an 
${\cal O}(\Delta \ln \sqrt{s})^2$ error.
If we want to obtain $\Psi$ from a finite difference approximation,
we need to interpolate $\Delta \ln \sqrt{s}\rightarrow 0$, but in this case the
experimental errors will most likely be much larger than the required
precision.  Such an interpolation procedure has already been applied in
ref. \cite{Neubert} near the $\tau$ region to test the running of
$\alpha_s$ (including  appropriate corrections to the leading twist
formalism). In this energy region the value of the QCD coupling is rather
large, and the interpolation yields evidence for some running. However, it
has also been pointed out in \cite{Neubert}, that the value of the
coupling extrapolated from the
$\tau$ region to high energies appears  small compared to direct
determinations.

\begin{figure}
\hspace*{3.cm}
\psfig{figure=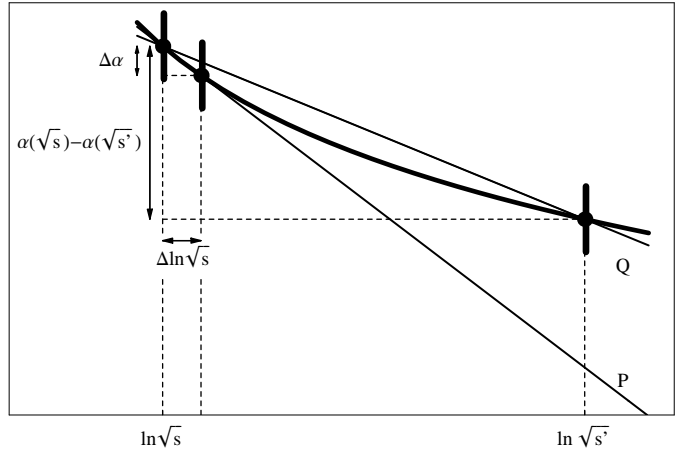,height=6cm}
\caption{\footnotesize Finite difference approximation of \(\Psi\).
If one takes  \(\Delta \ln \protect\sqrt{s}\protect\) very 
small, the errors can be
larger than \(\Delta \alpha\), and the result will be meaningless. This
can be avoided by choosing very far separated points 
\(\protect\sqrt{s}\protect\) and 
\(\protect\sqrt{s'}\protect\), but
then the approximation yields the slope of line \(Q\) instead of that of
\(P\).}
\end{figure}

In the next section we shall use the effective charge formalism to derive
several expressions within  leading twist QCD which relate the intrinsic
$\Psi_A$ function of $\alpha_A$
directly to the observables ${\cal O}_A$. We shall show that with just
three data points we can obtain good sensitivity to the value of $\Psi_0$
without any numerical differentiation or fit.

\subsubsection{Differential Commensurate Scale Relations}

Let us  formally differentiate eq.~(\ref{RAf}) with respect to $\Lambda$
\begin{eqnarray}
\frac{d{\cal O}_{A\,f}(\Lambda)}{d\Lambda}&=&\frac{2\,C}{\Lambda}
\left[\lambda_2^2f(\lambda_2)\,{\cal O}_A(\Lambda_2)-
\lambda_1^2f(\lambda_1)\,{\cal O}_A(\Lambda_1)\right]\nonumber\\
&-&\frac{2{\cal O}_{A\,f}(\Lambda)}{\Lambda}
                 -\frac{C}{\Lambda}
\int_{(\lambda_1\Lambda)^2}^{(\lambda_2 \Lambda)^2}
\frac{ds}{\Lambda^2}\,{\cal O}_A(\sqrt{s})
                 \frac{\sqrt{s}}{\Lambda}
\frac{df(\sqrt{s}/\Lambda)}{d(\sqrt{s}/\Lambda)}.
\label{dif1}
\end{eqnarray}
The first term in the right-hand side can be obtained directly
from the data on ${\cal O}_A$.
This is also the case for the second term,
after using eqs.~(\ref{RA}) and (\ref{alfaAk}), since
\begin{equation}
{\cal O}_{A\,f}(\Lambda)={\cal O}_{A\,f}^0
\left( \delta_A+ \frac{\alpha_{A\,f}(\Lambda)}{\pi}\right)
={\cal O}_{A\,f}^0\left( \delta_A+ \frac{\alpha_{A}(\Lambda_f)}{\pi}\right)
=\frac{{\cal O}_{A\,f}^0}{{\cal O}_A^0}\,{\cal O}_A(\Lambda_f),
\label{RaffromRA}
\end{equation}
Note that ${\cal O}_{A\,f}^0$
and ${\cal O}_A^0$ are known constants. Finally,
there is a choice of
$f(\xi)$ which allows us to recast
the third term in the right-hand side of eq.~(\ref{dif1})
and provide
a direct relation between the data and the effective charge.
Namely, we choose
\begin{equation}
\xi\,\frac{df(\xi)}{d\xi}=\rho\,f(\xi),
\label{dif2}
\end{equation}
with $\rho$ {\em any real number}. That is,
up to an irrelevant multiplicative constant, we take
\begin{equation}
f(\xi)=\xi^\rho.
\label{dif3}
\end{equation}

With this choice  eq.~(\ref{dif1}) can be simply written as
\begin{equation}
\frac{d{\cal O}_{A\,\rho}(\Lambda)}{d\Lambda}
=\frac{2C}{\Lambda}
\left[\lambda_2^{\rho+2}{\cal O}_A(\Lambda_2)-
\lambda_1^{\rho+2}{\cal O}_A(\Lambda_1)\right]-
\frac{\rho+2}{\Lambda}{\cal O}_{A\,\rho}(\Lambda).
\label{dif4}
\end{equation}
Note that, to simplify the notation, we have substituted the
$f$ subscript by $\rho$. In terms of $\Psi_A$ this means
\begin{equation}
\Psi_{A\,\rho}(\Lambda)=\Lambda\,
\frac{d\alpha_{A\,\rho}(\Lambda)}{d\Lambda}=\frac{\pi\Lambda}{{\cal
O}_{A\,\rho}}\,
\frac{d{\cal O}_{A\,\rho}(\Lambda)}{d\Lambda}.
\label{prebetaAf}
\end{equation}
But using its definition, we can easily see that
\begin{equation}
{\cal O}_{A\,\rho}^0=\frac{2\,C\,{\cal O}_A^0}{(\rho+2)}
(\lambda_2^{\rho+2}-\lambda_1^{\rho+2}),
\end{equation}
so that, using eq.~(\ref{RaffromRA}), we arrive at
\begin{equation}
\Psi_{A\,\rho}(\Lambda)=\pi\frac{\rho+2}{{\cal O}_{A}^0}
\left[\frac{\lambda_2^{\rho+2}{\cal O}_A(\Lambda_2)
-\lambda_1^{\rho+2}{\cal O}_A(\Lambda_1)}
{\lambda_2^{\rho+2}-\lambda_1^{\rho+2}}-{\cal O}_A(\Lambda_{\rho})
\right],
\label{betaAk}
\end{equation}
Note that we have just written
$\Psi_{A\,\rho}(\Lambda)$ directly in terms of observables.
Therefore, we have related the universal
$\Psi_0$ and $\Psi_1$ coefficients directly to observables,
without any dependence on the renormalization scheme or scale.

Up to this point $\Lambda_1$ and  $\Lambda_2$ are
arbitrary. In order to illustrate the meaning of eq.(\ref{betaAk}),
we now choose $\lambda_1=0$ and $\lambda_2=1$, so that
eq.(\ref{betaAk}) becomes:

\begin{equation}
\Psi_{A\,\rho}(\Lambda)=\pi\frac{\rho+2}{{\cal O}_{A}^0}
\left[{\cal O}_A(\Lambda)-{\cal O}_A(\Lambda_{\rho})
\right],
\label{betaL0L1}
\end{equation}

Let us remark that, although it may look similar, the above equation is 
not the finite difference
approximation
\begin{equation}
\Psi_{A}(\Lambda)\simeq\frac{\pi\,\Lambda}{{\cal O}^0_{A}} \,
\frac{{\cal O}_A(\Lambda)-{\cal O}_A(\Lambda-\Delta \Lambda)}{\Delta \Lambda} +
O(\Delta \Lambda^2)
\label{Numder}
\end{equation}
which is a good {\em numerical} approximation to $\Psi_A(\Lambda)$ when
$\Delta \Lambda$ is very small. In contrast, eq.~(\ref{betaAk}), is {\em exact}
(at leading twist) no matter whether $\Lambda-\Lambda_\rho$ is big or small.

However, we do not want to set $\lambda_1=0$, since then
 the integrated effective charges defined in
eq.(\ref{RAf}), contain higher twist contributions
which are unsuppressed at low energies, and our
leading twist formulae would be invalid in practice.
In addition, some observables like the number of
jets produced in $e^+e^-$ annihilation are only well defined
above some energy, which becomes
a lower cutoff in the integral of eq.(\ref{RAf}).

Nevertheless, by choosing $\Lambda$ and $\lambda_2$ appropriately,
we can obtain any value of $\Lambda_1\neq0$ and $\Lambda_2\neq0$,
even if we set $\lambda_1=1$, and so we will do so in the following.
That is:
\begin{equation}
\Psi_{A\,\rho}(\Lambda)=\pi \frac{\rho+2}{{\cal O}_{A}^0}
\left[\frac{\lambda_2^{\rho+2}{\cal O}_A(\Lambda_2)
-{\cal O}_A(\Lambda)}
{\lambda_2^{\rho+2}-1}-{\cal O}_A(\Lambda_\rho)
\right],
\label{betaL11}
\end{equation}
which is an exact formula relating $\Psi_A$
with the observable ${\cal O}_A$ at
three scales $\Lambda<\Lambda_\rho<\Lambda_2$.

It happens, however, that we are interested in measuring
 not the $\Psi_{A\,\rho}$
intrinsic function but $\Psi_A$ itself. We thus arrive at our
final result:
\begin{equation}
\Psi_{A}(\Lambda\,\lambda_\rho(\Lambda))\left[1+\frac{\lambda'_\rho}
{\lambda_\rho}
\right]=\pi\,\frac{\rho+2}{{\cal O}_{A}^0}
\left[\frac{\lambda_2^{\rho+2}{\cal O}_A(\Lambda_2)
-{\cal O}_A(\Lambda)}
{\lambda_2^{\rho+2}-1}-{\cal O}_A(\Lambda \lambda_\rho)
\right].
\label{betaA}
\end{equation}
where we have also defined $\lambda_\rho=\Lambda_\rho/\Lambda$.
Note that $\Psi_A$ appears in the above equation both
at $\Lambda_\rho$ and $\Lambda$ through
the $\lambda'_\rho$ coefficient, defined as $d\lambda/dLog\Lambda$,
which only vanishes at leading order.
Therefore, if we include higher order contributions
the above equation
is not enough to determine $\Psi_A$ at one given scale.

Let us work out first the implications of eq.(\ref{betaA})
at leading order, since it contains all the relevant features
of our approach.

\subsection{Leading order}

Suppose then that we had
three experimental data points at $s_a<s_b<s_c$.
In order to apply eq.~(\ref{betaL11}), we first identify
$\Lambda_2=\sqrt{s_c/s_a}$ and then
we obtain the $\rho$ such that $\sqrt{s_a}=\sqrt{s_b}/\lambda_\rho$.

The $I_{k\rho}$ integrals are given by
\begin{equation}
I_{k\rho}=\frac{k!}{\rho/2+1}\sum_{j=1}^2
\left[(-1)^j\,\lambda_2^{\rho+2}\sum_{l=0}^k\left(
\frac{(\ln\lambda_2^2)^{(k-l)}(-1)^l}{(\rho/2+1)^l(k-l)!}\right)
-\frac{(-1)^k}{(\rho/2+1)^k}\right].
\end{equation}
Thus, at leading order we have to obtain 
$\rho$ from
\begin{equation}
\ln\frac{s_b}{s_a}=2\ln\lambda_\rho=\frac{I_{1\rho}}{I_{0\rho}}=
\frac{s_c^{\rho/2+1}\ln(s_c/s_a)}{s_c^{\rho/2+1}-s_a^{\rho/2+1}}
-\frac{1}{\rho/2+1},
\label{k1}
\end{equation}
which can be evaluated numerically.

As we have already commented,
at leading order $\lambda'=0$, and therefore
\begin{equation}
\Psi_{A}(\sqrt{s_b})=\pi\,\frac{\rho+2}{{\cal O}_{A}^0}
\left[\frac{s_c^{\rho/2+1}{\cal O}_A(\sqrt{s_c})
-s_a^{\rho/2+1}{\cal O}_A(\sqrt{s_a})}
{s_c^{\rho/2+1}-s_a^{\rho/2+1}}-{\cal O}_A(\sqrt{s_b})
\right].
\label{betaLO}
\end{equation}

Let us remark once more that these are leading-twist formulae,
and $s_a,s_b,s_c$ should lie in a range where
higher twist effects are negligible.

\subsection{Beyond leading order}

As we have already seen, if we go beyond the leading order
contributions, we have to use eq.(\ref{betaA}), which
does not completely determine the value of $\Psi_A$ at a single scale.
In principle, we  need an additional equation.
In fact, the $\lambda'$ term can be neglected.
Intuitively, this is due to the very slow evolution of $\alpha_A$.
Let us give some numerical values; first, we will write
\begin{equation}
\frac{\lambda'_\rho}{\lambda_\rho}=\Psi_A(\Lambda)\Omega_\rho(\Lambda),
\end{equation}
with
\begin{equation}
\Omega_\rho(\Lambda)\equiv\frac{d\,\ln\lambda_\rho}{d(\alpha_A(\Lambda))}
=2\sum_{n=1}^\infty n\,a_\rho^{(n)}
\left(\frac{{\cal O}_A(\Lambda)}{{\cal O}_A^0}-\delta_A\right)^{n-1}.
\label{Omega}
\end{equation}
From PQCD we know that the expansion of $\Psi_A$  starts with
$\alpha_A^2$. Thus, the $\lambda'$ term in  eq.(\ref{betaA})
is an $O(\alpha_A^4)$ effect. It should
only be taken into account if we are
interested in $\Psi$ up to that order.
Numerically, the expected value of $\Psi_A(\Lambda)$
at the energies we will be using, ranges from
$10^{-2}$ to $2\times 10^{-2}$ at most.
In addition, $\Omega$ ranges
from  $3\times 10^{-2}$  to 0.5. Thus, even in the worst case,
the $\lambda'$ term contribution would be slightly
smaller than $1\%$ of
$\Psi$.
If that term is to be kept, then we need
and additional equation involving a fourth data point.
We have found that the final error estimate increases since it is much
harder to accommodate four points sufficiently separated within a
given energy range. It seems that  $1\%$ accuracy is the lower limit for
this method. If additional higher twist corrections are included, it
could be possible to extend the energy range to separate the points
and improve the precision.

Therefore, in what follows we will use eq.~(\ref{betaLO}).
However, the NLO $\rho$
parameter is now obtained by solving numerically the equation
\begin{eqnarray}
\ln\frac{s_b}{s_a}&=&
\frac{s_c^{\rho/2+1}\ln^2(s_b/s_a)}
{s_c^{\rho/2+1}-s_a^{\rho/2+1}}-\frac{1}{\rho/2+1}\\
&+&\frac{\Psi_0}{2}\left[\frac{(s_a\,s_c)^{\rho/2+1}\ln^2(s_b/s_a)}
{(s_c^{\rho/2+1}-s_a^{\rho/2+1})^2}-\frac{1}{(\rho/2+1)^2}
\right]\left(\frac{{\cal O}_A(\sqrt{s_a})}{{\cal O}^0_A}-\delta_A\right), \nonumber
\end{eqnarray}
where $s_a<s_b<s_c$ and $\sqrt{s_b}=\lambda_\rho\,\sqrt{s_a}$
and $\sqrt{s_c}=\lambda_2\,\sqrt{s_a}$.
Note that now $\Psi_0$ is an input, but the output is the
NLO $\Psi$ function.

\section{Error estimates}

Although they have inspired our approach,
observables with
$\delta_A\neq0$ are not well suited for our method,
because the relative error in ${\cal O}_A(E)$
becomes at least one order of magnitude larger for the
effective charge $\alpha_A(E)$. For example,
using the $e^+e^-$ hadronic ratio defined in Sect.1,
if we introduce a  $1\%$ error in $R_{e^+e^-}$, the error in $\alpha_R$
is $O(20\%)$ and we have to separate the data points
over five orders of magnitude to obtain $\Psi_R$ with a $10\%$ precision.
In practice, that renders the method useless.

The problem we have described is avoided if we use
an observable with $\delta_A=0$.
That is the case, for instance, of the $e^+e^-$ annihilation
in more than two jets,
$\sigma_{>2-\hbox{jets}}(s,y)=\sigma_{\hbox{tot}}-\sigma_{2\,\hbox{jets}}$,
where $y$
is used to define when two partons are unresolved
\cite{lampe}
(i.e. their invariant mass squared is less than $ys$). This
process does not occur in the parton model since it requires,
at least, one gluon. Note that $\Psi_0$ and $\Psi_1$ are independent of
$y$.

At LO we can work with exact results, but as soon as we
introduce higher orders, there is some degree of truncation
in the formulae. We have therefore first constructed  simulated data
following a model that corresponds to the exact LO equations.
Let us remark that these are models, not QCD. They are obtained by
the truncation of $\alpha_A$ at a given order. Thus, in principle, they
will have some different features from QCD, as for instance, some residual
scale dependence. In the real world this will not occur. However,
we have worked out these examples for illustrative purposes
to obtain a rough estimate of the errors.

\subsubsection{Leading order}

What we call the LO model is to use
\begin{equation}
\frac{\alpha_A(Q)}{\pi}=\frac{\alpha(M_Z)}{\pi}
-\frac{\Psi_0}{2}\ln\left(\frac{Q^2}{M_Z^2}\right)
\left(\frac{\alpha_A(M_Z)}{\pi}\right)^2,
\end{equation}
exactly. We have taken $\alpha_A(M_Z)$ as the reference value for simplicity.
Note, however, that the derivative of the above expression is
\begin{equation}
\Psi_A=-\frac{\Psi_0}{2}\left(\frac{\alpha_A(M_Z)}{\pi}\right)^2,
\end{equation}
which is a constant which differs by $O(\alpha/\pi)^3$ terms from the LO
PQCD result
\begin{equation}
  \Psi_A(Q)=-\frac{\Psi_0}{2}\left(\frac{\alpha_A(Q)}{\pi}\right)^2.
\end{equation}

In Table 1 we can see the estimates of the relative errors
in our determination of $\Psi_A$, which depend
on the different position of the data points, as well as
in their errors $\Delta{\cal O}_A$. Since the observable
vanishes in the parton model, the relative error in $\alpha_A$ is
exactly that of ${\cal O}_A$.

\begin{table}

\begin{center}
\begin{tabular}{|c|c|c|c|c|}
\hline
$\sqrt{s_a}$ (GeV) & $\sqrt{s_b}$ (GeV) & $\sqrt{s_c}$ (GeV)&
 $\Delta{\cal O}_A/{\cal O}_A$ & $\Delta\Psi_A/\Psi_A$\\
\hline \hline
30 & 100 & 300 & $1\,\%$ & $3\,\%$\\ \cline{4-5}
&&& $3\,\%$ & $9.1\,\%$\\ \hline
400& 640 & 1000 & $1\,\%$ & $6.2\,\%$\\ \cline{4-5}
&&& $3\,\%$ & $18.6\,\%$ \\ \hline
500& 875 & 1000 & $1\,\%$& $4.9\%$\\ \cline{4-5}
&&&$3\,\%$& $14.6\,\%$\\ \hline
\end{tabular}
\end{center}
\caption{\footnotesize
Estimated relative errors in the determination of $\Psi_0$
using the LO equations. We  assume the relative error
$\Delta{\cal O}_A/{\cal O}_A$ in the measurements of ${\cal O}_A$.
The estimates  correspond to an observable
with a vanishing parton model contribution ($\delta_A=0$)
such as $e^+e^-$ annihilation into more than two jets,
$\sigma_{>2-\hbox{jets}}$.}
\end{table}

The results in the table deserve some comments.
\begin{itemize}
\item First, the values of $\sqrt{s_a}$ and $\sqrt{s_c}$ have to be chosen
to maximize their distance, within a region of constant $N_F$. Thinking
in terms of $\sigma_{>2-\hbox{jets}}$,
they correspond either to the region
where both energies are sufficiently above
the b-quark pair threshold but still below $t\bar{t}$ production,
or both are above the $t\bar{t}$ pair threshold, in regions accessible at NLC.

\item Second, we have chosen the same relative error for  the
measurements at the three points. The intermediate energy $\sqrt{s_b}$ is
then tuned to minimize the error, which is
obtained assuming the three ${\cal O}_A$ measurements
are independent.
\end{itemize}

Let us remark once again that we have not used at any
moment the value of $\Psi_0$, which is obtained from the data
using this method. If we want to
use higher order contributions, using the value of $\Psi_0$ as an input,
we would obtain
information about higher order coefficients, like $\Psi_1$ if
we were to work at NLO.

\subsubsection{Beyond leading order}

The NLO model is now given by:
\begin{eqnarray}
\frac{\alpha_A(Q)}{\pi}&=&\frac{\alpha(M_Z)}{\pi}
-\frac{\Psi_0}{2}\ln\left(\frac{Q^2}{M_Z^2}\right)
\left(\frac{\alpha_A(M_Z)}{\pi}\right)^2\nonumber\\
&+&\frac{1}{4}\left[\Psi_0^2\ln^2\left(\frac{Q^2}{\Lambda}\right)
-2\,\Psi_1\ln\left(\frac{Q^2}{M_Z^2}\right)\right]
\left(\frac{\alpha_A(M_Z)}{\pi}\right)^3.
\end{eqnarray}
and therefore, we obtain
\begin{equation}
\Psi_A(Q)=-\frac{\Psi_0}{2}\left(\frac{\alpha_A(Q)}{\pi}\right)^2
-\frac{\Psi_1}{2}\left(\frac{\alpha_A(Q)}{\pi}\right)^3,
\label{psianlo}
\end{equation}
which  is the QCD NLO $\Psi_A$ result up to
$O(\alpha/\pi)^4$ terms.
In contrast with the LO case, obtaining $\rho$ now
requires some truncation of the formulae when
passing from 
eqs.~(\ref{aR2order}) and (\ref{RAf}) to eq.~(\ref{aRf2order}).
This is very interesting since we can thus obtain an estimate
of the theoretical error due to truncation, which will  be present
in the real case too.
It can be seen in Table 2 in the rows where $\Delta{\cal O}_A=0$,
and it is usually $O(1\%)$.

Again we have also  considered the experimental
$\Delta{\cal O}_A(E_i)$ uncertainties. The final error given
in the last column is
estimated assuming that the four experimental errors
and the one due to  truncation
are all independent. Note that when
passing from a $1\%$ experimental error to a $3\%$, the total
error is not multiplied by 3, since the truncation error
does not scale.

\begin{table}
\begin{center}
\begin{tabular}{|c|c|c|c|c|}
\hline
$\sqrt{s_a}$ (GeV) & $\sqrt{s_b}$ (GeV) & $\sqrt{s_c}$ (GeV)&
 $\Delta{\cal O}_A/{\cal O}_A$ & $\Delta\Psi_A/\Psi_A$\\
\hline \hline
&&&$0\%$&$2\%$\\\cline{4-5}
30 & 100 & 300& $1\,\%$ & $2.7\,\%$ \\ \cline{4-5}
&&& $3\,\%$ & $7.5\,\%$\\ \hline
&&&$0\%$&$.9\%$\\\cline{4-5}
400& 640 & 1000& $1\,\%$ & $10\,\%$ \\ \cline{4-5}
&&& $3\,\%$ & $29\,\%$ \\ \hline
&&&$0\%$&$1\%$\\\cline{4-5}
500& 875 & 1000 & $1\,\%$& $10\%$\\ \cline{4-5}
&&&$3\,\%$& $30\,\%$\\ \hline
\end{tabular}
\caption{\footnotesize Error estimates at NLO.}
\end{center}
\end{table}

The fact that we obtain larger errors in the NLO case
may seem surprising, but it is not. The reason is that the LO
is a very crude approximation of the $\Psi_A$ QCD scaling behavior.
In the LO model, the $\Psi$ function was a constant,
but in the NLO it changes with the energy scale, as it occurs 
in the realistic case. Indeed, the evolution of $\alpha_A$
at high energies becomes much slower so that the difference 
between $\alpha_A$ at two given points is smaller at NLO than at LO.
Hence, for the same relative errors, the relative uncertainties in the
NLO $\Psi$ function are much bigger. Of course, we expect the real 
data to show a behavior much closer to the NLO model.

\subsection{Using more than three points}

The advantage of fitting the data is that we can
reduce the errors by larger statistics. But that is also true
for our method. Up to now we have only used
three points of data, but in the realistic case we expect to
have several points at each energy range. It is then possible to form
many triplets of data points, one at low energies ($\sqrt{s_a}$),
another at intermediate energies ($\sqrt{s_b}$), and a last one in the
highest range ($\sqrt{s_b}$). Each one of these triplets will yield different
values and errors for $\Psi$, which can later be treated statistically,
thus decreasing the error estimates given in Table 2.

\section{Conclusions}

We have obtained an exact and very simple
relation between the Gell Mann-Low $\Psi$ function of
an effective charge of an observable and its integrals. These results are
renormalization-scheme and renormalization-scale independent. By
choosing specific weight functions, these relations can provide an
experimental determination of the PQCD
$\Psi$ function, thus testing the theory and
setting bounds on the properties of new particles that would
modify the expected QCD behavior.

We have shown that a good candidate for
this study is the $e^+e^-$ annihilation to more than two jets,
since it is a pure QCD process. Even within the simple leading-twist
formalism, which limits the applicability range,
we have found that with just three precise
measurements in present or presently planned accelerators, it could
be possible to determine the  $\Psi$ function without
making a QCD fit or any interpolation and numerical
differentiation of the data,
eliminating the specific uncertainties of these methods.
Thus we can obtain a determination of $\Psi$ with different
systematics. It also seems possible to extend the method and ideas,
to include  higher twist effects which will allow
the use of a wider range of energies. This could result
in an even more powerful set of tests of perturbative QCD.

\section*{Acknowledgments}

C.M.  and J.R.P. thank
the Theory Group at SLAC for their kind hospitality.
J.R.P. acknowledges the  Spanish
Ministerio de Educaci\'on y Cultura for financial support, as well as the Departamento de F\'\i sica de Part\'\i culas of the Universidade de Santiago de Compostela for its hospitality. We
also thank  M. Melles, J. Rathsman and N. Toumbas for helpful
conversations.

\end{document}